\documentclass[prd,byrevtex
,preprint
,tightenlines
,showkeys
,showpacs
]{revtex4}
\usepackage{amsmath,amssymb,amsxtra}
\newcommand{\al}{\alpha}
\newcommand{\del}{\delta}
\newcommand{\ep}{\epsilon}
\newcommand{\pa}{\partial}
\newcommand{\e}{\mathrm{e}}
\newcommand{\bs}[1]{\boldsymbol{#1}}

\newcommand{\bx}{\boldsymbol{x}}
\newcommand{\vg}{\varGamma}
\begin{document}
\title[Long-wavelength Cosmological Perturbation ]{Long-wavelength
  Cosmological Perturbation \\ 
 in the Universe with Multiple Perfect Fluids}
\author{Yasusada Nambu}
\email {nambu@allegro.phys.nagoya-u.ac.jp}
\author{Shin-ichi Ohokata}
%%%
\affiliation{Department of Physics, Graduate School of Science, Nagoya 
University, Chikusa, Nagoya 464-8602, Japan}
\preprint{DPNU02-10}
%%%%%%%%%%%%%%%%%%%
% v.0.0 2002/3/5
% v.0.5          3/13
% submit         4/19
% resubmit       6/14
\date{June 14, 2002}
%%%%%%%%%%%%%%%%%%%
\begin{abstract}
  We investigate the large scale cosmological perturbation in the
  Universe with multiple perfect fluids.  Using the long-wavelength
  approximation with Hamilton-Jacobi method, we derive the formula for
  the gauge invariant comoving curvature perturbation. As an
  application of our approach, we examine the large scale perturbation
  in a brane cosmology.
\end{abstract}
%%%%%%%%%%%%%%%%%%%
 \maketitle
%%%%%%%%%%%%%%%%%%%%%%%%%%%%%%%%%%%%%%%%%%%%%%%%%%%%%%%%%%%%%%%%%%%%%%%%%%%%%
\section{Introduction}

The analysis of the large-scale cosmological perturbation is important
issue to obtain the information on the initial density fluctuation
which  was generated during the era of the inflationary expansion of
the Universe. The  scales of cosmological interest such as
large-angle cosmic microwave background (CMB) have spent most of their
time far outside the Hubble radius and have re-entered only relatively
recently in the history of the Universe.

If we concentrate on only  the evolution of the large-scale
fluctuation, the long-wavelength  approximation (gradient expansion)
of Einstein's equation with the Hamilton-Jacobi (HJ)
method\cite{salopek92,salopek93, parry94,nambu98} is convenient to
obtain the solution of the fluctuation. Following this method, the
large-scale fluctuation is given by the lowest order of the spatial
gradient expansion of the Einstein equation and can be recognized as
the fluctuation of the spatially dependent integration constants
contained in the spatially inhomogeneous ``background'' solution. The
large-scale fluctuation of the metric and the matter field can be
obtained by differentiating the solution of the background Einstein
equation with respect to the constant of integration contained in it.
This approach provides an easier method to give the solution of the
long-wavelength perturbation and is useful especially for the Universe
with multiple scalar fields or multiple fluids.

In this paper, we consider the behavior of the long-wavelength
curvature perturbation. We aim to derive the formula for the gauge
invariant curvature perturbation  for the Einstein gravity  with
multiple fluids using the long-wavelength approximation with the HJ
method.  Contrary to the  case of multiple scalar
fields\cite{parry94,nambu98}, we can obtain explicit scale factor
dependence of the curvature perturbation.
As an application of our formalism, we consider the large-scale
fluctuation in the brane
cosmology\cite{maartens00,langlois00,langlois01}. The analysis of the
perturbation in the brane cosmology is complicated because the
perturbation equations do not form a closed system on the brane in
general. But for the large-scale scalar  type fluctuation, the
perturbation equations contain a closed form on the
brane\cite{maartens00,langlois01}, which may be solved without solving
for the bulk perturbations. Bulk effects produce a non-adiabatic mode
of fluctuation caused by the degrees of freedom of the Weyl
curvature. The Universe can be treated as the system with multi
component perfect fluids.

The plan of this paper is as follows. In Sec.~II, we introduce the
long-wavelength approximation of Einstein's equation with perfect
fluid. In Sec.~III, we derive the expression for the gauge invariant
curvature perturbation using the HJ method. We apply this method to
the brane cosmology in  Sec.~IV.   Sec.~V is devoted to summary. We
use units in which $c=\hbar=8\pi G=1$ throughout the paper.

%\newpage
%%%%%%%%%%%%%%%%%%%%%%%%%%%%%%%%%%%%%%%%%%%%%%%%%%%%%%%%%%%%%%%%%%%%%%%%%%%%
\section{Long-wavelength Approximation of the Einstein Gravity with
  Perfect Fluids }
We consider the Universe with two component perfect fluids of which
equation of state is given by $p^{(i)}=(\vg_i-1)\rho^{(i)}~(i=1,2)$
where $\vg_i$ is assumed to be constant. The form of the line element
is
%%%
\begin{equation}
 ds^2=-(N^2-N^iN_i)dt^2+2N_idtdx^i+\e^{2\al}d\bx^2,
\end{equation}
%%%
where $N$ is the lapse function, $N^i$ is the shift vector and $e^\al$
is the scale factor of the Universe.  We assume that each component of
the metric can have spatial dependence. By introducing the velocity
potential $\chi_i$, the Hamiltonian which derives the long-wavelength
dynamics of the Universe is\cite{salopek93}
%%%
\begin{align}
 &H_T=\int d^3x (N\mathcal{H}+N^i\mathcal{H}_i), \notag \\
 &\mathcal{H}=
-\frac{\e^{-3\al}}{12}P_{\al}^2+\e^{3\al}(\rho^{(1)}+\rho^{(2)}),\quad
  \rho^{(1)}=\left(\e^{-3\al}P_{\chi_1}\right)^{\vg_1},\quad
  \rho^{(2)}=\left(\e^{-3\al}P_{\chi_2}\right)^{\vg_2}\\
  &\mathcal{H}_i=-\frac{1}{3}P_{\al,i}-P_{\al}\al_{,i}-P_{\chi_1}\chi_{1,i}
-P_{\chi_2}\chi_{2,i}.\notag
\end{align}
%%%
The  three velocity of the fluid is given by
%%%
\begin{equation}
 v^{(i)}_j=-\frac{1}{\ep^{(i)}}\chi_{i,j}.
\end{equation}
%%%
$\ep^{(i)}$ is the specific enthalpy of the fluid which has the
following relation to the number density $n^{(i)}$:
%%%
\begin{equation}
\ep^{(i)}=\frac{\rho^{(i)}+p^{(i)}}{n^{(i)}},\quad n^{(i)}
=\left(\frac{\pa p^{(i)}}{\pa \ep^{(i)}}\right)_s.
\end{equation}
%%%
Using these relations, the specific enthalpy can be written as
%%%
\begin{equation}
\ep^{(i)}=\vg_i\left(\rho^{(i)}\right)^{\frac{\vg_i-1}{\vg_i}},
\end{equation}
%%%
and the three velocity is
%%%
\begin{equation}
v^{(i)}_j=-\e^{-3\al}\frac{P_{\chi_i}}{\vg_i\rho^{(i)}}\chi_{i,j}
\end{equation}
%%%
By variating this Hamiltonian with respect to $\al, P_{\al},\chi_i,
P_{\chi_i}, N, N^i$, we have the following equations
%%%
\begin{align}
 \dot\al&=\frac{\del{H}_T}{\del P_{\al}}=-\frac{N}{6}\e^{-3\al}P_{\al}
-\frac{1}{3}N^i{}_{,i}-N^i\al_{,i}, \\
 \dot\chi_i&=\frac{\del{H}_T}{\del P_{\chi_i}}
 =N\vg_i\e^{-3\vg_i\al}\left(P_{\chi_i}\right)^{\vg_i-1}-N^j\chi_{i,j}, \\
 -\dot
 P_{\al}&=\frac{\del{H}_T}{\del\al}
 =\frac{N}{4}\e^{-3\al}P_{\al}^2+3\e^{3\al}\left((\vg_1-1)\rho^{(1)}
 +(\vg_2-1)\rho^{(2)}\right)+(N^jP_{\al})_{,j},\\
 -\dot P_{\chi_i}&=\frac{\del{H}_T}{\del\chi_i}=(N^j\chi_i)_{,j}, \\
 \mathcal{H}&=0,\quad\mathcal{H}_i=0. \label{eq:HCMC}
\end{align}
%%%
Equations~\eqref{eq:HCMC} are the Hamiltonian constraint and the
momentum constraint.  If we assume that the shift vector is zero, we
have
%%%
\begin{align}
 &3\left(\frac{\dot\al}{N}\right)^2=\rho^{(1)}+\rho^{(2)}, \\
 &-\frac{2}{N}\left(\frac{\dot\al}{N}\right)\spdot
-3\left(\frac{\dot\al}{N}\right)^2=(\vg_1-1)\rho^{(1)}+(\vg_2-1)\rho^{(2)},\\
 &\dot\rho^{(1)}+3\dot\al\vg_1\rho^{(1)}=0,\quad\dot\rho^{(2)}
+3\dot\al\vg_2\rho^{(2)}=0
\end{align}
%%%
Although these equations have the  same form  as  for a flat
Friedman-Robertson-Walker (FRW) universe with perfect fluids,
variables $\al, \rho^{(1)}, \rho^{(2)}$ can have the spatial
dependence and this system represents the evolution of the Universe
with the long-wavelength inhomogeneity. The spatial dependence of each
variable is determined by the momentum constraint.

We treat this system using HJ method. By introducing the generating
functional $S[\al,\chi_1,\chi_2]$, the conjugate momenta of the
dynamical variables are replaced by
%%%
\begin{equation}
 P_\al=\frac{\del S}{\del\al},\quad P_{\chi_1}
=\frac{\del S}{\del\chi_1},\quad P_{\chi_2}=\frac{\del S}{\del\chi_2}. 
\end{equation}
%%%
The Hamiltonian constraint and the momentum constraint become
%%%
\begin{align}
 &\mathcal{H}=\e^{-3\al}
 \left[-\frac{1}{12}\left(\frac{\del S}{\del\al}\right)^2+\e^{6\al}
\left(\e^{-3\al}\frac{\del S}{\del\chi_1}\right)^{\vg_1}
 +\e^{6\al}\left(\e^{-3\al}\frac{\del
 S}{\del\chi_2}\right)^{\vg_2}\right]=0,
 \\
 &\mathcal{H}_i=\frac{1}{3}\left(\frac{\del S}{\del\al}\right)_{,i}
-\left(\frac{\del S}{\del\al}\right)\al_{,i}
-\left(\frac{\del S}{\del\chi_1}\right)\chi_{1,i}
-\left(\frac{\del S}{\del\chi_2}\right)\chi_{2,i}=0.
\end{align}
%%%
For  $N^i=0$, the evolution equations  are
%%%
\begin{align}
 &\dot\al=-\frac{N}{6}\e^{-6\al}\left(\frac{\del S}{\del\al}\right), \\
 &\dot\chi_1=N\vg_1\left(\e^{-3\al}\frac{\del
     S}{\del\chi_1}\right)^{\vg_1-1},
\quad
 \dot\chi_2=N\vg_2\left(\e^{-3\al}\frac{\del S}{\del\chi_2}\right)^{\vg_2-1}
\end{align}
%%%
By assuming the following form of the generating functional
%%%
\begin{equation}
 S=-2\int d^3x \e^{3\al}H(\chi_1,\chi_2),
\end{equation}
%%%
the Hamiltonian constraint becomes
%%%
\begin{equation}
\label{eq:HJ-hc}
 -3H^2+(-2H_{\chi_1})^{\vg_1}+(-2H_{\chi_2})^{\vg_2}=0.
\end{equation}
%%%
As this equation does not contain $\chi_1$ and $\chi_2$ explicitly,
the form of its solution can be written as
%%%
\begin{equation}
 H=H(\chi_1+d_1, \chi_2+d_2),
\end{equation}
%%%
where $d_1(\bx)$ and $d_2(\bx)$ are spatially dependent constants of
integration.  The momentum constraint becomes
%%%
$$
 \mathcal{H}_i=2\e^{3\al}\left[-H_{,i}+H_{\chi_1}\chi_{1,i}
 +H_{\chi_2}\chi_{2,i}\right]=-2\e^{3\al}(H_{d_1}d_{1,i}+H_{d_2}d_{2,i})=0.
$$
%%%
For $N^i=0$, the conjugate momentum $P_{\chi_1}$ and $P_{\chi_2}$
become constant in time and   we have
%%%
\begin{equation}
\label{eq:HJ-conserv}
 P_{\chi_1}=-2\e^{3\al}H_{\chi_1}=-2c_1,\quad
 P_{\chi_2}=-2\e^{3\al}H_{\chi_2}
 =-2c_2,
\end{equation}
%%%
where $c_1(\bx)$ and $c_2(\bx)$ are spatially dependent
constants. Hence the momentum constraint gives the following relation
between spatially dependent constants :
%%%
\begin{equation}
\label{eq:HJ-const}
 c_1d_{1,i}+c_2d_{2,i}=0.
\end{equation}
%%%
The evolution equations are 
%%%
\begin{align}
 & \dot\al=NH, \notag \\
 & \dot\chi_1=N\vg_1(-2H_{\chi_1})^{\vg_1-1}, \label{eq:HJ-evo}\\
 & \dot\chi_2=N\vg_2(-2H_{\chi_2})^{\vg_2-1}. \notag
\end{align}
%%%
\eqref{eq:HJ-hc},\eqref{eq:HJ-conserv}, \eqref{eq:HJ-const} and
\eqref{eq:HJ-evo} are basic equations to analyze the long-wavelength
dynamics of the Universe. The spatial inhomogeneity of the dynamical
variables $\al,\chi_1,\chi_2$ comes from the spatially dependent
constants $c_1(\bx), c_2(\bx), d_1(\bx),d_2(\bx)$. Solving HJ equation
is formally equivalent to obtaining the homogeneous background
solution. Once the background solution with constants of integration
is obtained, the spatial dependence of  the long-wavelength
fluctuation is determined through the momentum constraint.

%%%%%%%%%%%%%%%%%%%%%%%%%%%%%%%%%%%%%%%%%%%%%%%%%%%%%%%%%%%%%%%%%%%%
\section{Linear Perturbation about a Flat FRW Universe}
To obtain the curvature perturbation, we consider the linear
perturbation about the homogeneous  background. This means that
variables in the equations~\eqref{eq:HJ-hc},\eqref{eq:HJ-conserv},
\eqref{eq:HJ-const} and \eqref{eq:HJ-evo} are linearized as follows :
%%%
\begin{align*}
 &\al\rightarrow\al(t)+\del\al(t,\bx),\quad
 N\rightarrow 1+n(t,\bx),\quad
 \chi_i\rightarrow\chi_i(t)+\del\chi_i(t,\bx),\\
 &c_i\rightarrow c_i+\del c_i(\bx),\quad
 d_i\rightarrow d_i+\del d_i(\bx).
\end{align*}
%%%
The background equations are given by
%%%
\begin{align}
 &-3H^2+(-2H_{\chi_1})^{\vg_1}+(-2H_{\chi_2})^{\vg_2}=0, \notag \\
 &\dot\al=H, \quad \dot\chi_1=\vg_1(-2H_{\chi_1})^{\vg_1-1},\quad 
  \dot\chi_2=\vg_2(-2H_{\chi_2})^{\vg_2-1}. 
\end{align}
%%%
These equations describes the evolution of spatially flat FRW
universe. The constants contained in these equations do not have
spatial dependence.  By using equation~\eqref{eq:HJ-conserv}, the
background solution of the velocity potential can be written as
%%%
\begin{equation}
\label{eq:back-sol}
 \chi_1=\chi_1(\al,c_1,c_2,d_1,d_2),\quad \chi_2=\chi_2(\al,c_1,c_2,d_1,d_2).
\end{equation}
%%%
The evolution equations for linear perturbation are
%%%
\begin{align}
 & \del\dot\al=nH+H_{\chi_1}\del\chi_1+H_{\chi_2}\del\chi_2, \label{eq:HJmc} \\
 & \del\dot\chi_1=n\vg_1(-2H_{\chi_1})^{\vg_1-1} \notag \\
 &\qquad-2\vg_1(\vg_1-1)(-2H_{\chi_1})^{\vg_1-2}(H_{\chi_1\chi_1}\del\chi_1
 +H_{\chi_1\chi_2}\del\chi_2+H_{\chi_1d_1}\del d_1+H_{\chi_1d_2}\del
 d_2), 
 \label{eq:HJev1}\\
 & \del\dot\chi_2=n\vg_2(-2H_{\chi_2})^{\vg_2-1} \notag \\
 &\qquad
 -2\vg_2(\vg_2-1)(-2H_{\chi_2})^{\vg_2-2}(H_{\chi_1\chi_2}\del\chi_1
 +H_{\chi_2\chi_2}\del\chi_2+H_{\chi_2d_1}\del d_1+H_{\chi_2d_2}\del
 d_2).
 \label{eq:HJev2}
\end{align}
%%%
In equation~\eqref{eq:HJmc}, we do not have terms which contain $\del
d_i$ because this equation give the relation between $\del\al$ and
$\del\chi_i$, and is not the evolution equation. Indeed this equation
has the same form as the momentum constraint in the conventional
cosmological perturbation theory\cite{kodama84, mukhanov92}.

We solve the perturbation equation in $\del\al=0$ gauge. This gauge
condition corresponds to the zero curvature gauge\cite{hwang91}. In
this gauge, by eliminating the lapse function $n$ and using the
background solution \eqref{eq:back-sol}, we can rewrite the evolution
equation for $\del\chi_i$ as the following form:
%%%
\begin{equation}
\begin{pmatrix}\del\chi_1+\del d_1 \\ \del\chi_2+\del d_2 \end{pmatrix}_{,\al}=
\bs{X}_{,\al}\bs{X}^{-1}
\begin{pmatrix}\del\chi_1+\del d_1 \\ \del\chi_2+\del d_2 \end{pmatrix} 
+\frac{H_{\chi_1}\del d_1+H_{\chi_2}\del d_2}{H^2}
\begin{pmatrix}\vg_1(-2H_{\chi_1})^{\vg_1-1}\\
  \vg_2(-2H_{\chi_2})^{\vg_2-1}
\end{pmatrix}, \label{eq:evo-chi}
\end{equation}
%%%
where the matrix $\bs{X}$ is defined by using the background solution
\eqref{eq:back-sol} as follows
%%%
\begin{equation}
\bs{X}\equiv 
\begin{pmatrix}
 \chi_{1,c_1} & \chi_{1,c_2} \\ \chi_{2,c_1} & \chi_{2,c_2}
\end{pmatrix}. 
\end{equation}
%%%
If we substitute the following form of the solution to the evolution
equation \eqref{eq:evo-chi},
%%%
\begin{equation}
\begin{pmatrix}\del\chi_1+\del d_1 \\ \del\chi_2+\del d_2 \end{pmatrix}=\bs{X}
\begin{pmatrix} C_1(t) \\ C_2(t) \end{pmatrix},
\end{equation}
%%%
we have
%%%
\begin{equation}
 \bs{X}\begin{pmatrix}C_1 \\ C_2 \end{pmatrix}_{,\al}
 =c_0\frac{\e^{-3\al}}{H}\begin{pmatrix} \chi_{1,\al}\\ 
\chi_{2,\al}\end{pmatrix},
\end{equation}
%%%
where $c_0=c_1\del d_1+c_2\del d_2$ is a constant which has no spatial
dependence. This comes from equation~\eqref{eq:HJ-const}. Therefore,
$\del\chi_1$ and $\del\chi_2$ are given by
%%%
\begin{equation}
 \begin{pmatrix} \del\chi_1 \\ \del\chi_2 \end{pmatrix}=\bs{X}\left[
 \begin{pmatrix} \del c_1 \\ \del c_2 \end{pmatrix}
 +c_0\int
 d\al\frac{\e^{-3\al}}{H}\bs{X}^{-1}\begin{pmatrix}\chi_{1,\al} \\ 
\chi_{2,\al} \end{pmatrix}\right]
 -\begin{pmatrix} \del d_1 \\ \del d_2 \end{pmatrix},
\end{equation}
%%%
where $\del c_1(\bx)$ and $\del c_2(\bx)$ are spatially dependent constants.

Now we evaluate the gauge invariant variable corresponding to the
curvature perturbation on comoving slice
%%%
\begin{align}
 \mathcal{R}&\equiv-\del\al-H\frac{(\rho^{(1)}+p^{(1)})v^{(1)}+(\rho^{(2)}+p
 ^{(2)})v^{(2)}}{(\rho^{(1)}+p^{(1)})+(\rho^{(2)}+p^{(2)})} \notag \\
 &=-\del\al+\frac{H}{\dot
 H}(H_{\chi_1}\del\chi_1+H_{\chi_2}\del\chi_2), 
\label{eq:R1}
\end{align}
%%%
where we have used equation (6) to express the velocity perturbation
by $\delta\chi$. 
This quantity gives the spatial curvature perturbation on the comoving
slice and equivalent to Bardeen's parameter $\zeta$ in the
long-wavelength limit. To proceed our calculation, we prepare formulas
which comes from the conservation law
equation~\eqref{eq:HJ-conserv}. By differentiating
equation~\eqref{eq:HJ-conserv} with respect to $\al$, we have
%%%
\begin{equation}
 \bs{M}\begin{pmatrix}\chi_{1,\al} \\ \chi_{2,\al}\end{pmatrix}=-3
 \begin{pmatrix} H_{\chi_1} \\ H_{\chi_2} \end{pmatrix},\quad \bs{M}\equiv
 \begin{pmatrix} H_{\chi_1\chi_1} & H_{\chi_1\chi_2} \\
   H_{\chi_1\chi_2} 
& H_{\chi_2\chi_2} \end{pmatrix}.
\end{equation}
%%%
By differentiating equation~\eqref{eq:HJ-conserv} with respect to
$c_1, c_2$, we have
%%%
$$
 \bs{M}\bs{X}=\e^{-3\al}\begin{pmatrix} 1 & 0 \\ 0 & 1 \end{pmatrix}.
$$
%%%
Using these relations, the gauge invariant curvature perturbation
$\mathcal{R}$ in the $\del\al=0$ gauge becomes
%%%
\begin{align}
\mathcal{R}&=\frac{H}{\dot
  H}(H_{\chi_1}\del\chi_1+H_{\chi_2}\del\chi_2) 
\notag \\
 &=-\frac{1}{3}\frac{\frac{\del c_1}{c_1}\vg_1\rho^{(1)}
+\frac{\del c_2}{c_2}\vg_2\rho^{(2)}}{\vg_1\rho^{(1)}+\vg_2\rho^{(2)}}
+c_0\left(\int dt\e^{-3\al}-\frac{H}{\dot H}\e^{-3\al}\right)
\end{align}
%%%
where energy density of each fluids is given by
$\rho^{(1)}=(-2c_1)^{\vg_1}\e^{-3\vg_1\al},
\rho^{(2)}=(-2c_2)^{\vg_2}\e^{-3\vg_2\al}$.  This is the main result
of this paper. The first term of this expression corresponds to the
contribution of the growing mode of adiabatic and iso-curvature
perturbation. This term is same as the one derived by using the local
conservation of energy momentum tensor\cite{wands00}. The second term
corresponds to decaying mode. In our analysis, the decaying mode
cannot have spatial dependence, but it gives the correct time
dependence. For $\vg_1=\vg_2$ , the system reduces to a single
component fluid case and $\mathcal{R}$ obeys the following evolution
equation
%%%
\begin{alignat}{2}
 &\dot{\mathcal{R}}=0 & \quad&\text{for}\quad \vg_1=1, \\
 &\ddot{\mathcal{R}}+3\dot\al\dot{\mathcal{R}}=0& &\text{for}\quad\vg_1\ne 1
\end{alignat}
%%%

In the situation when the energy density of one fluid dominates
$\rho^{(1)}\gg\rho^{(2)}$, the time dependence of the scale factor is
$\e^{\al}\approx (t/t_0)^\frac{2}{3\vg_1}$ and the behavior  of the
curvature perturbation is given by
%%%
\begin{align}
 \mathcal{R}&\approx -\frac{\del c_1}{c_1}
-\frac{1}{3}\left(\frac{\del c_2}{c_2}
-\frac{\del c_1}{c_1}\right)\frac{\vg_2\rho_2}{\vg_1\rho_1}
+c_0\left(\int dt\e^{-3\al}-\frac{H}{\dot H}\e^{-3\al}\right) \notag \\
 &=-\frac{\del c_1}{c_1}-\frac{1}{3}\left(\frac{\del c_2}{c_2}
-\frac{\del
  c_1}{c_1}\right)\frac{\vg_2(-2c_2)^{\vg_2}}{\vg_1(-2c_1)^{\vg_1}}
\times \e^{-3(\vg_2-\vg_1)\al}-c_0 t_0
\times \frac{2(\vg_1-1)}{2-\vg_1}\e^{\frac{3}{2}(\vg_1-2)\al}.
\end{align}
%%%
The first term corresponds to the adiabatic growing mode which is constant
in time. The second term corresponds to the iso-curvature mode and the
third term is the decaying mode. The decaying mode vanishes in the
$\vg_1=1$ case.
%%%%%%%%%%%%%%%%%%%%%%%%%%%%%%%%%%%%%%%%%%%%%%%%%%%%%%%%%%%%%%%%%
\section{Application to the Brane Cosmology}
We apply the method developed in Sec.~III to evaluate the large scale
curvature perturbation in the brane
cosmology\cite{maartens00,langlois00,langlois01}.  By assuming the
Randall-Sundrum type brane model, the four dimensional Einstein
equation on the  spatially homogeneous brane universe is
%%%
\begin{align}
 &3H^2=\rho+\frac{l^2}{12}\rho^2+\rho_{\mathcal{E}} ,\notag\\
 &-2\dot H-3H^2=p+\frac{l^2}{12}(\rho^2+2p\rho)+\frac{\rho_{\mathcal{E}}}{3},
 \label{eq:brane-frw} \\
 &\dot\rho+3H(\rho+p)=0,\quad p=(\vg-1)\rho,\quad
 \dot\rho_{\mathcal{E}}
+4H\rho_{\mathcal{E}}=0, \notag
\end{align}
%%%
where $l^2$ is the inverse of brane tension and $\rho_{\mathcal{E}}$ is
the energy density of the Weyl matter. The unconventional feature of
the brane cosmology is the appearance of $\rho^2$  and
$\rho_{\mathcal{E}}$ terms on the right hand side of the Einstein
equation. The Hamiltonian for the long-wavelength dynamics which
derives  equation~\eqref{eq:brane-frw} in the homogeneous limit is
%%%
\begin{align}
 &{H}_T=\int d^3x (N\mathcal{H}+N^i\mathcal{H}_i), \\
 &\mathcal{H}=-\frac{\e^{-3\al}}{12}P_{\al}^2
 +\e^{3\al}\left(\rho+\frac{l^2}{12}\rho^2+\rho_{\mathcal{E}}\right),
 \quad\rho=\left(\e^{-3\al}P_{\chi_1}\right)^\vg,\quad\rho_{\mathcal{E}}
 =\left(\e^{-3\al}P_{\chi_2}\right)^{\frac{4}{3}}, \\
 &\mathcal{H}_i=\frac{P_{\al,i}}{3}-P_\al\al_{,i}
-P_{\chi_1}\chi_{1,i}-P_{\chi_2}\chi_{2,i}.
\end{align}
%%%
The effective energy density and the effective pressure for the
perfect fluid are defined by
%%%
\begin{align}
 &\rho_{\mathcal{M}}=\rho+\frac{l^2}{12}\rho^2, \\
 &p_{\mathcal{M}}=(\vg-1)\rho+\frac{l^2}{12}(2\vg-1)\rho^2.
\end{align}
%%%
The specific enthalpy for the effective matter field is given by
%%%
$$
 \ep_{\mathcal{M}}
 =\vg\rho^{\frac{\vg-1}{\vg}}\left(1+\frac{l^2}{6}\rho^2\right). 
$$
%%%
For the Weyl matter,
%%%
$$
 \ep_{\mathcal{E}}=\frac{4}{3}\rho_{\mathcal{E}}{}^{1/4}.
$$
%%%
Hence the three velocity of the effective matter field and the Weyl
matter becomes
%%%
\begin{align*}
 &v_{\mathcal{M}}=-\frac{\del\chi_1}{\ep_{\mathcal{E}}}
 =-\frac{\del\chi_1}{\vg\rho^{\frac{\vg-1}{\vg}}
\left(1+\frac{l^2}{6}\rho^2\right)}, \\
 &v_{\mathcal{E}}=-\frac{\del\chi_2}{\ep_{\mathcal{E}}}
 =-\frac{\del\chi_2}{\frac{4}{3}\rho_{\mathcal{E}}^{1/4}}.
\end{align*}
%%%
The gauge invariant variable corresponding to the curvature
perturbation on comoving slice is
%%%
\begin{align}
 \mathcal{R}&=-\del\al-H\frac{(\rho_{\mathcal{M}}
 +p_{\mathcal{M}})v_{\mathcal{M}}+(\rho_{\mathcal{E}}
 +p_{\mathcal{E}})v_{\mathcal{E}}}{(\rho_{\mathcal{M}}+p_{\mathcal{M}})
 +(\rho_{\mathcal{E}}+p_{\mathcal{E}})} \notag \\
 &=-\del\al+H\frac{\rho^{\frac{1}{\vg}}\del\chi_1
 +\rho_{\mathcal{E}}{}^{3/4}\del\chi_2}{\vg\left(\rho
 +\frac{l^2}{6}\rho^2\right)+\frac{4}{3}\rho_{\mathcal{E}}} \notag \\
 &=-\del\al+\frac{H}{\dot H}(H_1\del\chi_1+H_2\del\chi_2).
\end{align}
%%%
This expression is same as the result of standard cosmology
\eqref{eq:R1}. The explicit form obtained in the $\del\al=0$ gauge is
%%%
\begin{equation}
 \mathcal{R}=-\frac{1}{3}\frac{\vg\left(\rho+\frac{l^2}{6}\rho^2\right)
 \left(\frac{\del c_1}{c_1}\right)+\frac{4}{3}
 \rho_{\mathcal{E}}\left(\frac{\del c_2}{c_2}\right)}{\vg\left(\rho
 +\frac{l^2}{6}\rho^2\right)+\frac{4}{3}\rho_{\mathcal{E}}}
 +c_0\left(\int dt\e^{-3\al}-\frac{H}{\dot H}\e^{-3\al}\right).
\end{equation}
%%%
The first term of this expression reproduces the result of
Ref.~\cite{langlois00} which was derived by using the local
conservation law of the fluids. Our method include the contribution of
the decaying mode.  The  time dependence of the curvature perturbation
can be obtained in the following three cases:
%%%%%%%%%%%%%%%%%%%
\paragraph{$l^2\rho^2$ dominates}
The time dependence of the scale factor is
$\e^{\al}\approx(t/t_0)^{\frac{1}{3\vg}}$ and
%%%
\begin{align}
 \mathcal{R}&\approx -\frac{\del c_1}{3c_1}
 -\frac{8}{3\vg}\left(\frac{\del c_2}{c_2}
 -\frac{\del c_1}{c_1}\right)\frac{\rho_{\mathcal{E}}}{l^2\rho^2}
 +c_0\left(\int dt\e^{-3\al}-\frac{H}{\dot H}\e^{-3\al}\right) \notag \\
 &=-\frac{\del c_1}{3c_1}-\frac{8}{3\vg}\left(\frac{\del c_2}{c_2}
 -\frac{\del c_1}{c_1}\right)\frac{(-2c_2)^{4/3}}{l^2(-2c_1)^{2\vg}}
 \times \e^{2(3\vg-2)\al}-c_0t_0\times \frac{2\vg-1}{1-\vg}\e^{3(\vg-1)\al}.
\end{align}
%%%%%%%%%%%%%%%%%%%%
\paragraph{$\rho_{\mathcal{E}}$ dominates}
The time dependence of the scale factor is $\e^{\al}\approx (t/t_0)^{1/2}$ and
%%%
\begin{align}
 \mathcal{R}&\approx -\frac{\del c_2}{3c_2}
 -\frac{\vg}{4}\left(\frac{\del c_2}{c_2}
 -\frac{\del c_1}{c_1}\right)\left(\frac{\rho}{\rho_{\mathcal{E}}}
 +\frac{l^2}{6}\frac{\rho^2}{\rho_{\mathcal{E}}}\right)
 +c_0\left(\int dt\e^{-3\al}-\frac{H}{\dot H}\e^{-3\al}\right) \notag \\
 &=-\frac{\del c_2}{3c_2}-\frac{\vg}{4}\left(\frac{\del c_2}{c_2}
 -\frac{\del c_1}{c_1}\right)\left(\frac{(-2c_1)^{\vg}}{(-2c_2)^{4/3}}
 \times\e^{(4-3\vg)\al}+\frac{l^2}{6}\frac{(-2c_1)^{2\vg}}{(-2c_2)^{4/3}}
\times \e^{-2(3\vg-2)\al}\right)\\
 &\qquad\qquad\qquad\qquad -c_0t_0\times \e^{-\al}. \notag
\end{align}
%%%%%%%%%%%%%%%%%%%%
\paragraph{$\rho$ dominates}
The time dependence of the scale factor is $\e^{\al}\approx
(t/t_0)^{\frac{2}{3\vg}}$ and
%%%
\begin{align}
 \mathcal{R}&\approx -\frac{\del c_1}{3c_1}-\frac{4}{9\vg}
 \left(\frac{\del c_2}{c_2}-\frac{\del c_1}{c_1}\right)
 \frac{\rho_{\mathcal{E}}}{\rho}+c_0\left(\int dt\e^{-3\al}
-\frac{H}{\dot H}\e^{-3\al}\right) \notag \\
 &=-\frac{\del c_1}{3c_1}-\frac{4}{9\vg}\left(\frac{\del c_2}{c_2}
 -\frac{\del c_1}{c_1}\right)\frac{(-2c_2)^{4/3}}{(-2c_1)^{\vg}}
 \times \e^{-(4-3\vg)}-c_0t_0\times\frac{2(\vg-1)}{2-\vg}
\e^{-\frac{3}{2}(2-\vg)\al}.
\end{align}

%%%%%%%%%%%%%%%%%%%%%%%%%%%%%%%%%%%%%%%%%%%%%%%%%%%%%%%%%%%%%%%%%
\section{Summary}
We have derived a formula of long-wavelength curvature perturbation
for the Universe with multi-component fluids using HJ method.  The
non-decaying mode of the curvature perturbation is explicitly
expressed as the function of  scale factor of the Universe. This  mode
is the mixture of the adiabatic growing mode which is constant in time
and the time varying iso-curvature mode. In the case of the
multiple-scalar field\cite{nambu98}, the curvature perturbation cannot
be expressed explicitly as the function of the scale factor  (see
Appendix) because we cannot obtain the solution of the background
scalar fields as the function of the scale factor in general. For the
perfect fluids case, each component of fluids evolves separately and
this makes it possible to obtain the background  and the perturbation
solution.

Although the decaying mode obtained in this paper reproduces the same
time dependence  as  the result of  the conventional cosmological
perturbation theory, it cannot have the spatial dependence because of
the momentum constraint. The homogeneous decaying mode obtained here
can be absorbed to the background   solution by re-defining the
background constant of integration. This is related to the feature of
decaying mode in zero curvature gauge and co-moving gauge ; the
decaying mode under these gauge conditions becomes higher order in
long-wavelength expansion compared with that of  other gauge
conditions\cite{hwang02}.

As the  application of the HJ method, the large scale curvature
perturbation in the brane cosmology was derived. Our method reproduced
the result obtained by using standard cosmological perturbation theory
with the local conservation of the energy-momentum
tensor\cite{langlois00}.  It is interesting to investigate the brane
model with  the scalar field. But  we could not  obtain the
Hamiltonian which derives the long-wavelength dynamics of the brane
cosmology with the scalar field.  This is left as our future problem.

%\begin{acknowledgments}
%The author would like to thank ............
%\end{acknowledgments}

%%%%%%%%%%%%%%%%%%%%%%%%%%%%%%%%%%%%%%%%%%%%%%%%%%%%%%%%%%%%%%%%%%
\appendix
\section{Curvature Perturbation in the Multiple Scalar Fields System}
We review the derivation of the long-wavelength curvature perturbation
in the Einstein gravity with multiple scalar fields\cite{nambu98}. HJ
equation and the evolution equation are
%%%
%%
\begin{align}
 &-3H^2+2H_{\phi_1}^2+2H_{\phi_2}^2+V(\phi_1,\phi_2)=0, \label{eq:sc-hj} \\
 &\dot\al=NH, \label{eq:sc-al}\\
 &\dot\phi_1=-2NH_{\phi_1},\quad \dot\phi_2=-2NH_{\phi_2} \label{eq:sc-phi}
\end{align}
%%%
The solution of the HJ equation is given by
$H=H(\phi_1,\phi_2,d_1,d_2)$ where $d_1(\bx)$ and $d_2(\bx)$ are
spatially dependent constants. From the momentum constraint, the
spatial dependence of these constants are determined:
%%%
$$
 H_{d_1}d_{1,i}+H_{d_2}d_{2,i}=0.
$$
%%%
Other constants of motion in the system are given by
%%%
\begin{equation}
 \e^{3\al}H_{d_1}=c_1(\bx)\equiv \e^{-3\al_0(\bx)},\quad 
\e^{3\al}H_{d_2}=c_2(\bx)\equiv \e^{-3\al_0(\bx)}f(\bx) , \label{eq:sc-cons}
\end{equation}
%%%
From this, we can express $\phi_1, \phi_2$ as the function of the scale factor:
%%%
\begin{equation}
 \phi_1=\phi_1(\al+\al_0, f,d_1,d_2),\quad \phi_2=\phi_2(\al+\al_0, f,d_1,d_2).
\end{equation}
%%%
The evolution equation for the perturbation is
%%%
\begin{align}
 &\del\dot\al=nH+H_{\phi_1}\del\phi_1+H_{\phi_2}\del\phi_2, \\
 &\del\dot\phi_1=-2nH_{\phi_1}-2(H_{\phi_1\phi_1}\del\phi_1
 +H_{\phi_1\phi_2}\del\phi_2+H_{\phi_1d_1}\del d_1+H_{\phi_1d_2}\del d_2), \\
 &\del\dot\phi_2=-2nH_{\phi_2}-2(H_{\phi_1\phi_2}\del\phi_1
 +H_{\phi_2\phi_2}\del\phi_2+H_{\phi_2d_1}\del d_1+H_{\phi_2d_2}\del d_2).
\end{align}
%%%
The gauge invariant variable corresponding to the curvature
perturbation on comoving slice  is given by
%%%
\begin{equation}
 \mathcal{R}=-\del\al-\frac{H^3}{2H_{\al}}(\phi_{1,\al}\del\phi_1
 +\phi_{2,\al}\del\phi_2).
\end{equation}
%%%
We evaluate this expression in the $\del\al=0$ gauge. For this
purpose, we prepare the relation  derived form
equation~\eqref{eq:sc-cons}. By differentiating
equation~\eqref{eq:sc-cons} with respect to $c_1, c_2$, we have
%%%
\begin{equation}
\bs{M}\bs{X}=\e^{-3\al}\bs{1},\quad\bs{M}=
\begin{pmatrix}H_{\phi_1d_1} & H_{\phi_2d_1}\\ H_{\phi_1d_2} 
& H_{\phi_2d_2} \end{pmatrix},\quad\bs{X}=
\begin{pmatrix}\phi_{1,c_1} & \phi_{1,c_2}\\ \phi_{2,c_1} &
  \phi_{2,c_2} 
\end{pmatrix}
\end{equation}
%%%
The evolution equation in the $\del\al=0$ gauge can be written as
%%%
\begin{equation}
\begin{pmatrix}\del\phi_1 \\ \del\phi_2 \end{pmatrix}_{,\al} 
=\bs{X}_{,\al}\bs{X}^{-1}
\begin{pmatrix}\del\phi_1 \\ \del\phi_2 \end{pmatrix}
-\frac{2}{H}
\begin{pmatrix} H_{\phi_1d_1} & H_{\phi_1d_2} 
\\ H_{\phi_2d_1} & H_{2d_2} \end{pmatrix}
\begin{pmatrix} \del d_1 \\ \del d_2 \end{pmatrix}. \label{eq:zeroc-evo}
\end{equation}
%%%
By substituting 
%%%
$$
\begin{pmatrix}\del\phi_1 \\ \del\phi_2 \end{pmatrix}
=\bs{X}
\begin{pmatrix} C_1(t) \\  C_2(t)\end{pmatrix}
$$
%%%
to the evolution equation, we have
%%%
$$
\bs{X}\begin{pmatrix} C_1 \\  C_2 \end{pmatrix}_{,\al}=
-\frac{2}{H}\bs{M}^T\begin{pmatrix}\del d_1 \\ \del d_2 \end{pmatrix}=
-\frac{2\e^{-3\al}}{H}(\bs{X}^{-1})^T\begin{pmatrix}\del d_1 
\\ \del d_2 \end{pmatrix}
$$
%%%
and the solution of equation~\eqref{eq:zeroc-evo} is given by
%%%
\begin{equation}
\begin{pmatrix}\del\phi_1 \\ \del\phi_2 \end{pmatrix}=\bs{X}\left[
\begin{pmatrix} \del c_1 \\  \del c_2 \end{pmatrix}
+\int d\al\frac{\e^{-3\al}}{H}(\bs{X}^{-1})(\bs{X}^{-1})^T
\begin{pmatrix}\del d_1 \\ \del d_2 \end{pmatrix}
\right],
\end{equation}
%%%
where $\del c_1, \del c_2$ are spatially dependent constants. Using
this solution, the curvature perturbation becomes
%%%
\begin{equation}
 \mathcal{R}=-\frac{H^3}{2H_{\al}}
 \begin{pmatrix}\phi_{1,\al} & \phi_{2,\al} \end{pmatrix}
 \bs{X}\left[
\begin{pmatrix} \del c_1 \\  \del c_2 \end{pmatrix}
 +\int d\al\frac{\e^{-3\al}}{H}(\bs{X}^{-1})(\bs{X}^{-1})^T
\begin{pmatrix}\del d_1 \\ \del d_2 \end{pmatrix}
\right].
\end{equation}
%%%
For the growing mode $\del d_1=\del d_2=0$, we have
%%%
\begin{equation}
 \mathcal{R}=\del\al_0+\frac{\phi_{1,\al}\phi_{1,f}
+\phi_{2,\al}\phi_{2,f}}{(\phi_{1,\al})^2+(\phi_{2,\al})^2}\del f.
\end{equation}
%%%
The first term corresponds to the adiabatic growing mode which is
constant in time, and the second term corresponds to the iso-curvature
mode. The explicit scale factor dependence can be obtained with the
assumption of the slow rolling approximation and  the specific form of
the scalar field potential\cite{nambu98}.

%%%%%%%%%%%%%%%%
%\section*{References}
%\bibliography{paper}

\end{document}